
\input phyzzx
\hoffset=0.2truein
\voffset=0.1truein
\hsize=6truein
\def\TITLEPAGE{\frontpagetrue}
\def\CALT#1{\hbox to\hsize{\tenpoint \baselineskip=12pt
	\hfil\vtop{\hbox{\strut CALT-68-#1}
	\hbox{\strut DOE RESEARCH AND}
	\hbox{\strut DEVELOPMENT REPORT}}}}

\def\CALTECH{\smallskip
	\address{California Institute of Technology, Pasadena, CA 91125}}
\def\TITLE#1{\vskip 1in \centerline{\fourteenpoint #1}}
\def\AUTHOR#1{\vskip .5in \centerline{#1}}

\def\ENDTITLEPAGE{\vfil\eject\pageno=1}

\def\sqr#1#2{{\vcenter{\hrule height.#2pt
      \hbox{\vrule width.#2pt height#1pt \kern#1pt
        \vrule width.#2pt}
      \hrule height.#2pt}}}

\def\section#1#2{
\noindent\hbox{\hbox{\bf #1}\hskip 10pt\vtop{\hsize=5in
\baselineskip=12pt \noindent \bf #2 \hfil}\hfil}
\medskip}

\def\underwig#1{	
	\setbox0=\hbox{\rm \strut}
	\hbox to 0pt{$#1$\hss} \lower \ht0 \hbox{\rm \char'176}}

\def\bunderwig#1{	
	\setbox0=\hbox{\rm \strut}
	\hbox to 1.5pt{$#1$\hss} \lower 12.8pt
	 \hbox{\seventeenrm \char'176}\hbox to 2pt{\hfil}}

\TITLEPAGE
\CALT{1740}
\TITLE{Superstring Compactification and Target Space Duality
\foot{Work supported in part by the U.S. Department of
Energy under Contract No. DE-AC0381-ER40050}}
\AUTHOR{John H. Schwarz}

\CALTECH

\bigskip
\singlespace
\centerline{ABSTRACT}
\smallskip

{\narrower\narrower\tenpoint
This review talk focusses on some of
the interesting developments in the
area of superstring compactification that have occurred in the last
couple of years. These include the discovery that ``mirror symmetric"
pairs of Calabi--Yau spaces, with completely distinct geometries and
topologies, correspond to a single (2,2) conformal field theory. Also,
the concept of target-space duality, originally discovered for toroidal
compactification, is being extended to Calabi--Yau spaces. It also
associates sets of geometrically distinct manifolds to a single conformal
field theory.

A couple of other topics are presented very briefly. One concerns
conceptual challenges in reconciling gravity and quantum mechanics. It
is suggested that certain ``distasteful allegations" associated with
quantum gravity such as loss of quantum coherence,
unpredictability of fundamental parameters of particle physics, and
paradoxical features of black holes are likely to be circumvented by
string theory. Finally there is a brief discussion of the importance of
supersymmetry at the TeV scale, both from a practical point of view and
as a potentially significant prediction of string theory. \par}

\bigskip

\noindent Presented at Strings and Symmetries 1991 \hfill May 1991
\ENDTITLEPAGE
\vfill\eject

\normalspace

\noindent {\bf 1. Introduction}

\REF\jhsworld{J. H. Schwarz, {\it Int. J. Mod. Phys.}
{\bf A2} (1987) 593--643;
{\it Int. J. Mod. Phys.} {\bf A4} (1989) 2653--2713.}

The conference organizers have asked me to give a review survey of significant
developments in superstring compactification that have
occurred in the last couple of years since the last review papers
that I wrote on this subject.${}^{\jhsworld}$
A great deal of impressive progress has been
made, and it will only be possible to survey a portion of it. The
choice of topics is based mostly on what has caught my attention, and what
I have been able to digest. There are undoubtedly many important developments
that will not be mentioned.

The two main topics to be discussed are Calabi--Yau compactification
and target-space duality. Two important developments in Calabi--Yau
compactification will be stressed. The first is the existence of
holomorphic prepotentials that determine the K\"ahler potentials
that describe the moduli spaces ${\cal M}_{21}$ and ${\cal M}_{11}$, associated
with complex structure deformations and K\"ahler form deformations,
respectively. The second is the
remarkable mirror symmetry that associates a pair of Calabi--Yau spaces
to the same (2,2) conformal field theory. They are related to one
another by interchange of the moduli spaces ${\cal M}_{21}$ and ${\cal
M}_{11}$.

The most famous example of target--space duality is the $R\to 1/R$
symmetry associated with compactification on a circle of radius $R$. As
with mirror symmetry this transformation relates distinct geometries
that are associated with the same conformal field theory. Generalizations
appropriate to toroidal compactifications have been known for some time and
will
be reviewed very briefly. The little bit that is known about
such symmetries in the case of K3 and Calabi--Yau space compactification
will also be discussed. An interesting proposal has been made to
restrict the possibilities for low-energy effective actions that
incorporate nonperturbative supersymmetry breaking
by the duality symmetries.

\REF\jhscurse{J. H. Schwarz, ``Target Space Duality
and the Curse of the Wormhole,"
CALT-68-1688, to be published in {\it Beyond the Standard Model II}, the
proceedings of a conference held at the University of Oklahoma (World
Scientific 1991).}

Recently, my interest in target-space duality was reactivated
by the realization that it can be viewed as a discrete symmetry group
that is a subgroup of spontaneously broken continuous gauge
symmetries---what
has been referred to as `local gauge symmetry.' This
led me to propose that it could have a bearing on resolving certain
deep problems associated with quantum gravity.${}^{\jhscurse}$
After reviewing some of
the disturbing allegations that are made about the inevitable consequences
of reconciling general relativity and quantum mechanics,  I will discuss
possible ways in which they may be circumvented in string theory.

The concluding section makes a plea to demonstrate to our experimental
colleagues that our work is relevant. It is suggested that a
strong case could be made for supersymmetry at the TeV scale as an almost
inevitable feature of any quasi-realistic string vacuum. This being so,
perhaps it would not be inappropriate for us to stick our necks out a bit
and call this a `prediction of string theory.'

\medskip

\noindent{\bf 2. Progress in Calabi--Yau Compactification}

Let me begin by recalling a few basic facts. By definition, a Calabi--Yau space
is
a K\"ahler manifold of three (complex) dimensions and vanishing
first Chern class. The complex (Dolbeault) cohomology groups $H^{p,q}$ have
dimensions $b_{p q}$ given by the Hodge diamond
$$
\matrix{&&&1\cr
&&0&&0\cr
&0&&b_{11} & &0\cr
1&&b_{21} &&b_{21} &&1\cr
&0&&b_{11} & &0\cr
&&0& &0\cr
&&&1}$$

One generator of $H^{1,1}$ is the K\"ahler form
$J=i g_{\mu \bar\nu} dx^\mu \wedge dx^{\bar \nu}$. Thus  $b_{11} \geq 1$. A
Calabi--Yau space, with specified K\"ahler class,
admits a unique Ricci-flat metric. Also, there exists a covariantly
constant spinor $\lambda$, in terms of which the holomorphic three-form is
$\Omega_{\mu\nu\rho}=\lambda\gamma_{\mu\nu\rho}\lambda$.

In the context
of heterotic string compactification,
the existence of $\lambda$ is responsible for
the fact that the 4D low-energy theory has N=1 supersymmetry.
Altogether, the
massless spectrum contains the following
N=1 supermultiplets:

a) N=1 supergravity (graviton and gravitino)

b) Yang--Mills supermultiplets (adjoint vectors and spinors for $E_6\times
E_8\times\dots$)

c) Various chiral supermultiplets (Weyl spinor and a scalar)

The chiral supermultiplets include matter and moduli multiplets. The matter
multiplets consist of $b_{11}$ generations (27 of $E_6$) and
$b_{21}$ antigenerations ($\overline{27}$ of $E_6$). Which of these
one chooses to call
`generations' and `antigenerations' is purely a matter
of convention, of course.
The moduli consist of an `S field' and `T fields.' The S field
contains the dilaton $\phi$ and the axion $\theta$. The vev of the dilaton
gives the string coupling constant ($<\phi>\sim 1/g^2$), and $\theta$ is
the 4D dual of the antisymmetric tensor field $B_{\mu\nu}$. Locally, the
vevs of $\phi$ and $\theta$ parametrize the coset manifold $SU(1,1)/U(1)$.

The T fields consist of $b_{11}$ $E_6$ singlets whose vevs parametrize the
moduli space of K\"ahler form deformations, ${\cal M}_{11}$, and $b_{21}$
$E_6$
singlets whose vevs parametrize the space of complex structure deformations,
${\cal M}_{21}$. Altogether, the Calabi--Yau moduli space is
the tensor product ${\cal M}_{CY}={\cal M}_{11}\times {\cal M}_{21}$.
Locally, this is the same thing as the space of (2,2)
conformal field theories, though the K\"ahler geometry of ${\cal
M}_{21}$ differs in the two cases by the effects of world sheet
instantons. However, as we will
discuss, they differ globally by duality symmetries, so that the moduli
space of (2,2) theories is given by
${\cal M}_{(2,2)}={\cal M}_{CY}/G$, where $G$ is
a discrete group. This is the space that classifies inequivalent string
compactifications.

\REF\fands{S. Ferrara and A. Strominger, in {\it Strings 89}, Proc.
Superstring Workshop, Texas A\&M Univ., ed. R. Arnowitt et al., (World
Scientific 1990); P. Candelas and X. C. de la Ossa, {\it Nucl. Phys.} {\bf
B355} (1991) 455.}

Remarkably, the moduli spaces ${\cal M}_{11}$ and ${\cal M}_{21}$ are
themselves
K\"ahler manifolds. Here I will simply state the salient facts without
attempting to give the proofs, which can be found in the
literature.${}^{\fands}$

Let $z^{\alpha}$, $\alpha=1,2,\dots,b_{21}$ be local complex coordinates for
${\cal M}_{21}$. Then linearly independent generators of $H^{2,1}$ are
given by
$$\chi_{\alpha\kappa\lambda\bar\mu}=-{1\over 2}\Omega_{\kappa\lambda\rho}
g^{\rho\bar\eta}{\partial\over\partial z^{\alpha}}g_{\bar\mu\bar\eta},
\quad \alpha=1,2,\dots,b_{21}.$$
Out of $\chi_{\alpha}=\chi_{\alpha\kappa\lambda\bar\mu}dx^{\kappa}\wedge
dx^{\lambda}\wedge d{\bar x}^{\bar\mu}$ and $\bar\chi_{\bar\beta}$ one
constructs the K\"ahler metric for ${\cal M}_{21}$ by the formula
$$G_{\alpha\bar\beta}  = - {\int_M \chi_{\alpha}\wedge\bar\chi_{\bar\beta}
\over \int_M \Omega\wedge\bar\Omega} = - {\partial\over\partial z^{\alpha}}
{\partial\over\partial {\bar z}^{\bar\beta}}
\log \left(i\int_M \Omega\wedge\bar\Omega\right) .$$
{}From this it follows that $K = - \log \left(i\int_M
\Omega\wedge\bar\Omega\right)$
is the K\"ahler potential.

Moreover, ${\cal M}_{21}$ is a K\"ahler manifold with a {\it holomorphic
prepotential}. (Such K\"ahler manifolds are sometimes said to be of `restricted
type.')
The formula is given most succinctly using projective
coordinates (in $P_{b_{21}}$) $z^a$, $a=1,2,\dots, b_{21}+1$, defined with
respect to a canonical homology basis of $H_3$. The basic cycles $A^a$ and
$B_b$ are arranged to intersect in a manner analogous to the $A$ and $B$
cycles of a Riemann surface. The coordinates $z^a$ are given by
$z^a = \int_{A^a}\Omega$ and the derivatives of the holomorphic prepotential
${\cal G}(z^a)$ are given by ${\partial\over\partial z^a} {\cal G} =
\int_{B_a}\Omega$. In terms of ${\cal G}$ and its complex conjugate
$\bar {\cal G}({\bar z}^{a})$ the K\"ahler potential is given by
$$ e^{-K} = -i\left( z^a{\partial\bar {\cal G}\over\partial {\bar z}^a}
-{\bar z}^a{\partial {\cal G}\over\partial {z}^a}\right) .$$
The prepotential also encodes the Yukawa couplings of the
antigenerations, which are given by its third derivatives.
(I am confident that more details of this construction will be presented by
other
speakers.)

The moduli space ${\cal M}_{11}$ has a very similar description. In terms of
local complex coordinates $w^A$, with $A=1,2,\dots,b_{11}$. Its metric
is given in terms of a K\"ahler potential by $G_{A\bar B} =
{\partial\over \partial w^A} {\partial\over \partial{\bar w }^{\bar B}} K$,
as usual. The K\"ahler potential is given by $K= - \log \kappa(J,J,J)$,
where $\kappa(J,J,J) = {4\over3} \int_M J\wedge J\wedge J$ and $J$ is
the K\"ahler form. In terms of
projective coordinates $w^i$, $i=1,2,\dots,b_{11}+1$,
there is again a holomorphic prepotential ${\cal F}(w^i)$ with
$$e^{-K} = -i\left(w^i {\partial \bar {\cal F}\over \partial {\bar w}^i} -
{\bar w}^i {\partial {\cal F} \over \partial w^i} \right).$$
As before, the third derivatives of ${\cal F}$ give the Yukawa couplings of the
generations. The significant asymmetry between ${\cal F}$ and ${\cal G}$ in the
case of
the geometric limit is that ${\cal F}$ is a cubic function so that the Yukawa
couplings are constants, whereas ${\cal G}$ is a complicated nonpolynomial
expression, so that those Yukawa couplings depend on the complex structure
deformation moduli.

A Calabi--Yau space corresponds to a geometric limit of a (2,2)
superconformal field theory. This geometric limit includes effects to
every order in $\alpha'$ in the associated sigma model, but it does
not include the nonperturbative effects associated with world-sheet
instantons. It turns out that these world-sheet instantons contribute
to the moduli space of K\"ahler form deformations ${\cal M}_{11}$ but
not to the moduli space of complex structure deformations ${\cal M}_{21}$.
As a result the prepotential ${\cal G}$ can be computed exactly in the
geometric
limit, but the prepotential ${\cal F}$ cannot. The earlier assertion that
${\cal F}$ is
a cubic function referred to the geometric limit. When the instanton
contributions
are included it is no longer cubic. It is the latter expression that is
relevant to
string theory. This modification
of Calabi--Yau geometry implied by the corresponding conformal field
theory is sometimes referred to as ``quantum geometry" in the recent
literature.

We now turn to the mirror symmetry conjecture. The similarity in the
description
of the two moduli spaces ${\cal M}_{11}$ and ${\cal M}_{21}$
suggests that to every Calabi--Yau space $M$ (with $b_{21} >0$) there is
a mirror partner Calabi--Yau space $\tilde M$ such that the two
moduli spaces are interchanged:
$$ \tilde {\cal M}_{11} = {\cal M}_{21}\quad {\rm and}\quad
\tilde {\cal M}_{21} = {\cal M}_{11}. $$
Clearly, in view of the remarks made above, this is only possible if
we include the instanton corrections in the description of the moduli
spaces. Thus the precise statement of the conjecture is that there is a
a mirror partner Calabi--Yau space such that  the two spaces correspond
to distinct geometric limits of the same (2,2) conformal field theory.
This a remarkable conjecture from a mathematical point of view, since
the CY spaces have completely different geometries and topologies. It
is also of some practical importance from the physical point of view,
since computations of the prepotentials are significantly easier to
carry out in the geometric limit (using topological formulas) than for the
exact CFT in general. By
computing moduli spaces of complex-structure deformations for a pair
of mirror CY spaces one deduces (by the equations above) the exact
geometry of the moduli spaces of K\"ahler form deformations, including
the effects of world-sheet instantons.

\REF\lerche{W. Lerche, C. Vafa, and N. Warner, {\it Nucl. Phys.} {\bf B324}
(1989) 427.}
\REF\candelas{P. Candelas, M. Lynker, and R. Schimmrigk,
{\it Nucl. Phys.} {\bf B341} (1990) 383.}

A considerable amount of evidence in support of the
mirror symmetry conjecture
has been amassed. I am not completely sure of the history, but I
believe the idea originated when Dixon noticed the
symmetrical way in which generations and anti-generations are treated in
Calabi-Yau compactification of the heterotic string \`a la Gepner, and
when Lerche, Vafa, and Warner noticed the symmetrical appearance of the
(c,c) and (a,c) rings in N=2 Landau--Ginzburg theory.${}^{\lerche}$
Candelas, Lynker,
and Schimmrigk computed the Hodge numbers $b_{11}$ and $b_{21}$ for
several thousand CY spaces that can be described as intersections of
weighted complex projective spaces and observed that there was an almost
perfect correspondence between pairs of CY spaces with these numbers
interchanged.${}^{\candelas}$
Some unpaired examples are to be expected in their list,
since it is certainly not complete.
Given that fact, it is remarkable how few of them
there are. Also, one class of spaces is certainly special.
There exist Calabi--Yau spaces with $b_{21}=0$. The mirror of such a space
should have $b_{11}=0$, but such a space cannot be a Calabi--Yau space, since
they always have the K\"ahler form itself as at least one generator
of $H^{1,1}$. The significance of this class of exceptions is being
investigated by the experts. It may be of some practical importance if
we hope to eventually find an example that gives three generations and no
antigenerations, so as to avoid the problem of understanding how extra
generations and anti-generations pair up to acquire a large mass. Such
a space with $b_{11}=3$ and $b_{21}=0$ might not exist, however.

\REF\greene{ B. R. Greene and M. R. Plesser, {\it Nucl. Phys.} {\bf B338}
(1990) 15.}
\REF\aspinwall{P. S. Aspinwall, C. A. L\"utken, and G. G. Ross, {\it Phys.
Lett.}
{\bf 241B} (1990) 373; P. S. Aspinwall and C. A. L\"utken, {\it Nucl. Phys.}
{\bf 355} (1991) 482.}
\REF\ferrara{S. Ferrara, M. Bodner, and A. C. Cadavid,
{\it Phys. Lett.} {247B} (1990) 25; {\it Class. Quant. Grav.}
{\bf 8} (1991) 789.}
\REF\delaossa{P. Candelas, X. C. de la Ossa, P. S. Green, and L. Parkes,
{\it Nucl. Phys.} {\bf B359} (1991) 21; {\it Phys. Lett.}
{\bf 258B} (1991) 118.}

Further evidence in support of the mirror symmetry conjecture and
insight into various structural details have been obtained from a
variety of additional studies.
These include analysis of explicit examples based
on (2,2) orbifolds${}^{\greene}$ and
additional studies of the Landau-Ginzburg
connection.${}^{\aspinwall}$ Another approach utilizes Gepner's correspondence
between Calabi--Yau
compactification and compactification of Type II superstring theories.
The constructions based on Type IIA and Type IIB theories turn out
to be related by mirror symmetry. This fact provides a rather powerful
tool for detailed studies, since it benefits from the restrictions
implied by the additional supersymmetry of the Type II
theories.${}^{\ferrara}$
Finally, I should mention the detailed investigation of a specific
mirror pair of CY spaces by Candelas {\it et al.}${}^{\delaossa}$
They compute the
prepotential ${\cal F}$ for the
one-dimensional moduli space of the K\"ahler class for the CY space
given by a quintic polynomial in $P_4$ and compare it to the prepotential
${\cal G}$ for the mirror space (given by orbifolding the original space
by a suitable discrete symmetry group). By comparing the two expressions
they are able to explicitly identify the instanton contributions and
infer the number of ``holomorphic
curves" of various sorts---deep results in algebraic geometry.

\medskip

\REF\green{M. B. Green, J. H. Schwarz, and L. Brink, {\it
Nucl. Phys.} {\bf B198} (1982) 474.}
\REF\kikkawa{K. Kikkawa and M. Yamasaki, {\it Phys. Lett.} {\bf 149B} (1984)
357; N. Sakai and I. Senda, {\it Prog. Theor. Phys.} {\bf 75} (1984) 692.}
\REF\alvarez{E. Alvarez and M. A. R. Osorio,
{\it Phys. Rev.} {\bf D40} (1989) 1150; D. Gross and I. Klebanov, {\it
Nucl. Phys.}
{\bf B344} (1990) 475; E. Smith and J. Polchinski, ``Duality Survives
Time Dependence," Univ. of Texas preprint
UTTG-07-91, Jan. 1991; A. A. Tseytlin, {\it Mod. Phys. Lett.} {\bf A16} (1991)
1721.}
\noindent{\bf 3. Target Space Duality}

The notion of target space duality is becoming a more and more prevalent
theme in string theory, with a wide range of applications and implications.
Target-space dualities are discrete symmetries of compactified string
theories, whose existence suggests a breakdown of geometric concepts at
the Planck scale. The simplest example is given by closed bosonic
strings with one dimension of space taken to form a circle of radius $R$.
In this case the momentum component of a string corresponding to this
dimension is quantized: $p=n/R$, $n\in Z$. This is a general consequence
of quantum mechanics and is not special to strings, of course. What is
special for strings is the existence of winding modes. A closed string
can wrap $m$ times around the circular dimension. A string state with
momentum and winding quantum numbers $n$ and $m$, respectively, receives
a zero-mode contribution to its mass-squared given by${}^{\green}$
$$M_0^2=(n/R)^2 + (mR/\alpha')^2, $$
where $\alpha'$ is the usual Regge slope parameter. It is evident
that the simultaneous interchanges $R\leftrightarrow
\alpha'/R$ and $m\leftrightarrow n$ leaves the mass formula
invariant.${}^{\kikkawa}$
In fact, the entire physics of the interacting theory is left unaltered
provided that one simultaneously rescales the dilaton field (whose
expectation value controls the string coupling constant) according to
$\phi \rightarrow \phi - ln(R/\sqrt{\alpha'})$.${}^{\alvarez}$
The basic idea is that
the radii $R$ and $\alpha'/R$ both correspond to the same ($c=1$)
conformal field theory, and it is the conformal field theory that
determines the physics. This is the simplest example of the general
phenomenon called ``target-space duality."

\REF\narain{K. S. Narain, {\it Phys. Lett.} {\bf 169B} (1986) 41; K. S.
Narain, M. H. Sarmadi, and E. Witten, {\it Nucl. Phys.}
{\bf B279} (1987) 369.}
\REF\shapere{V. P. Nair, A. Shapere, A. Strominger, and F. Wilczek,
{\it Nucl. Phys.} {\bf B287} (1987) 402;
A. Shapere and F. Wilczek, {\it Nucl. Phys.} {\bf B320} (1989)
669; A. Giveon, E. Rabinovici, and G. Veneziano, {\it Nucl. Phys.} {\bf B322}
(1989) 167.}
\REF\giveon{S. Ferrara, D. L\"ust, A. Shapere, and S. Theisen, {\it Phys.
Lett.} {\bf 225B} (1989) 363;
A. Giveon and M. Porrati, {\it Phys. Lett.} {\bf 246B} (1990) 54; {\it
Nucl. Phys.} {\bf B355} (1991) 422.}
The circular compactification described above has been generalized to
the case of a $d$-dimensional torus, characterized by $d^2$ constants
$G_{ab}$ and $B_{ab}$.${}^{\narain}$
The symmetric matrix $G$ is the metric of
the torus, while $B$ is an antisymmetric matrix.
These parameters describe the moduli space of
toroidal compactification and can be
interpreted as the vacuum expectation values of $d^2$ massless scalar
fields. The dynamics of the
toroidal string coordinates $X^a$ is described by the world sheet action
$$S=\int d^2\sigma[G_{ab}\partial^{\alpha}X^a\partial_{\alpha}X^b +
\epsilon ^{\alpha \beta}B_{ab} \partial_{\alpha}X^a\partial_{\beta}X^b]$$
In this case we can introduce $d$-component vectors of integers $m^a$
and $n_a$ to describe the winding modes and discrete momenta,
respectively. A straightforward calculation then gives the zero-mode
contributions
$$M_0^2=G_{ab}m^am^b+G^{ab}(n_a-B_{ac}m^c)(n_b-B_{bd}m^d)$$
where $G^{ab}$ represents the inverse of the matrix $G_{ab}$. The
one-dimensional case is recovered by setting $G_{11}=R^2/\alpha'$ and
$B=0$. The generalization of the duality symmetry becomes
$G+B\rightarrow (G+B)^{-1}$ and $\phi\rightarrow \phi - {1\over 2} ln
\det(G+B)$. The $B$ term in the world-sheet action is topological. The
parameters $B_{ab}$ are analogous to the $\theta$ parameter in QCD, and
the quantum theory is invariant under integer shifts $B_{ab}\rightarrow
B_{ab}+N_{ab}$. Combined with the inversion symmetry, these generate the
infinite discrete group $O(d,d;Z)$.${}^{\shapere}$ (The analogous group in
the case of the heterotic string is $O(d+16,d;Z)$.)
Thus, whereas the moduli $G_{ab}$
and $B_{ab}$ locally parametrize the coset space $O(d,d)/[O(d)\times
O(d)]$, points in this space related by $O(d,d;Z)$ transformations
correspond to the same conformal field theory and should be identified.
It is conjectured that when nonperturbative effects break the flatness of
the effective potential, so that the scalar fields that correspond to
the moduli can acquire mass, the
discrete duality symmetries of the theory are preserved.${}^{\giveon}$

\REF\dine{M. Dine, P. Huet, and N. Seiberg,
{\it Nucl. Phys.} {\bf B322} (1989) 301.}
The $O(d,d;Z)$ target space duality symmetries are discrete
remnants of spontaneously broken gauge symmetries.${}^{\dine}$
Specifically, for
values of the moduli corresponding to a fixed point of a subgroup of
this discrete group, the corresponding string background has enhanced
gauge symmetry. (This is most easily demonstrated by showing that there
are additional massless vector string states in the spectrum.) At such a
point some of the duality symmetry transformations coincide with finite gauge
transformations. By considering all possible such fixed points it is
possible to identify an infinite number of distinct gauge symmetries,
with all but a finite number of them spontaneously broken for
any particular choice of the moduli.

\REF\seiberg{N. Seiberg, {\it Nucl. Phys.} {\bf B303} (1988) 286.}
\REF\west{M. B. Green, J. H. Schwarz, and P. C. West, {\it Nucl. Phys.} {\bf
B254} (1985) 327.}
One may wonder whether the occurrence of target-space dualities is
special to toroidal compactification or whether it occurs generically
for curved compactification spaces such as Calabi--Yau manifolds. The
four-dimensional analog, namely K3 compactification, has been analyzed
in some detail.${}^{\seiberg}$ In that case (applied to the heterotic
string) the moduli space is $O(20,4)/[O(20)\times
O(4)]$, parametrized by 80 massless scalar fields and the
duality group is $O(20,4;Z)$. Remarkably, this is exactly the
same manifold and duality group that arises in the case of toroidal
compactification of the heterotic string to six dimensions. One might be
tempted to speculate that the two compactifications are equivalent,
but that cannot be correct since K3 compactification breaks half of
the supersymmetry while toroidal compactification does not break
any.${}^{\west}$ By modding out certain
symmetries of the torus, it is possible to form an orbifold
for which half the supersymmetry is broken and
the moduli space is still essentially the same. This orbifold seems
likely to correspond (at least locally)
to the same conformal field theory as K3.
Having spaces of distinct topology correspond to identical
conformal field theories goes beyond what we learned from tori. (There
various different geometries all having the same topology were
identified.) However, as we have seen, such
identifications do exist for Calabi--Yau spaces, which occur in
mirror pairs of opposite Euler number.

The moduli space of a
Calabi--Yau compactification factorizes into the manifold ${\cal M}_{21}$
that describes complex-structure deformations times the manifold
${\cal M}_{11}$ that describes K\"ahler form deformations.
There are duality symmetry transformations that act on
each of these spaces separately. The two classes of transformations
would seem to have very different
interpretations from a geometrical point of view. However, the two
factors are interchanged for the mirror manifold, so if one
associates them with a mirror pair of Calabi--Yau spaces,
then they appear on an equivalent footing. The problem is to determine
the target-space duality group that acts on each of these moduli spaces.
A natural action of the discrete group $Sp(2+2b_{21},Z)$ can be
defined on the third cohomology group
$$H^3 = H^{(3,0)}\oplus H^{(2,1)}\oplus H^{(1,2)}\oplus H^{(0,3)}$$
analogous to the symplectic modular group for Riemann surfaces.
This symplectic group contains the possible discrete symmetries of
${\cal M}_{21}$. The
mirror symmetry implies a corresponding action of $Sp(2+2b_{11},Z)$
on the space
$$H^{(0,0)}\oplus H^{(1,1)}\oplus H^{(2,2)}\oplus H^{(3,3)}$$
describing possible discrete symmetries of ${\cal M}_{11}$. Thus
altogether the target space duality of the (quantum corrected)
Calabi--Yau space should be given by some subgroup
$$G_{TD} \subseteq Sp(2+2b_{21},Z)\times Sp(2+2b_{11},Z).$$
Examples have been worked out in special cases.

\REF\font{A. Font, L. E. Ib\'a\~nez, D. L\"ust, and F. Quevedo, {\it Phys.
Lett.} {\bf 245B} (1990) 401; S. Ferrara, N. Magnoli, T. R. Taylor, and
G. Veneziano, {\it Phys. Lett.} {\bf 245B} (1990) 409; H. P. Nilles and M.
Olechowski, {\it Phys. Lett.} {\bf 248B} (1990) 268; P. Binetruy and M. K.
Gaillard, {\it Phys. Lett.} {\bf 253B} (1991) 119.}
\REF\cvetic{M. Cvetic, A. Font, L. E. Ib\'a\~nez, D. L\"ust, and F.
Quevedo, {\it Nucl. Phys.} {\bf B361} (1991) 194.}
\REF\lust{S. Ferrara, C. Kounnas, D. L\"ust, and F. Zwirner,
``Duality-Invariant Partition Functions and Automorphic Superpotentials
for (2,2) String Compactifications," preprint CERN-TH.6090/91; D.
L\"ust, ``Duality Invariant Effective String Actions and Automorphic
Functions for (2,2) String Compactifications," preprint CERN-TH.6143/91.}

A potentially important application of the duality symmetries has been
proposed in connection with the construction of low-energy effective
actions. The idea is that these should be exact symmetries of the
complete quantum theory and should still be present even after
nonperturbative effects (such as those that break supersymmetry) are
taken into account and after heavy fields are integrated out. This
means that, in terms of a low-energy effective action in four dimensions
with N=1 supersymmetry, the duality symmetries should be realized on
the superpotential, which is therefore restricted to be a suitable
automorphic function. This is a very significant restriction on the
characterization of the low-energy theory. Therefore there is some
hope for saying quite a bit about nonperturbative effects without solving
the difficult problem of computing them from first principles.
There is some evidence that the combination
of gluino condensation and duality symmetry are sufficient to remove  all
flat directions from the potential.${}^{\font}$
This means that the size of the compact space,
which is one of the moduli, is dynamically determined and
all the other parameters that determine the vacuum configuration are
also determined. Supersymmetry is broken and the
cosmological constant typically comes out negative (corresponding to
anti de Sitter space). However, there are some examples for which
the cosmological constant vanishes.${}^{\cvetic}$
Recently, there have been studies
of automorphic prepotentials for general (2,2)
compactifications.${}^{\lust}$

\medskip
\noindent{\bf 4. Conceptual Challenges in Reconciling Gravity and Quantum
Mechanics}

There are a variety of technical and conceptual obstacles that need to
be overcome if a satisfactory understanding of the reconciliation of
general relativity and quantum mechanics is to be achieved. These can be
divided into two categories---amazing requirements and distasteful
allegations. If string theory is the correct approach to constructing a
fully consistent unification of all fundamental forces, then it should
hold the keys to the right answers. In this case, our job is to discern
the clever tricks that string theory employs. This may sound like a
strange way to approach the problem, but string theory has proved to be
a fruitful source of inspiration in the past.
Examples range from the discovery of supersymmetry to unexpected anomaly
and divergence cancellation mechanisms and much more. It could hold
many more surprises in store for us.

One ``amazing requirement" is perturbative finiteness (or
renormalizability). This is not yet fully established, but there is
considerable evidence that this is achieved in string theory, even
though it is apparently impossible for any point-particle field theory
that incorporates general relativity (in four dimensions). A second
requirement is that causality have a precise meaning when the
space-time metric is a dynamical quantum field. This undoubtedly happens
in string theory, but it would be nice to understand in
detail just what is involved. Third, the theory should be applicable to
the entire universe, perhaps describing it by a single wave function.
An obvious question in this connection is whether string theory suggests
some special choice of boundary condition, such as that proposed by Hartle and
Hawking, and whether this could provide a rationale for selecting a
particular vacuum configuration.

\REF\mixed{S. W. Hawking, {\it Phys. Rev.} {\bf D14} (1976) 2460.}
\REF\worm{S. Coleman, {\it Nucl. Phys.} {\bf B307} (1988) 867; S.B. Giddings
and A. Strominger, {\it Nucl. Phys.} {\bf B307} (1988) 854.}
\REF\prob{S. Coleman, {\it Nucl. Phys.} {\bf B310} (1988) 643; B. Grinstein
and M. Wise, {\it Phys. Lett.} {\bf 212B} (1988) 407; J. Preskill, {\it
Nucl. Phys.}
{\bf B323} (1989) 141; S. W. Hawking, {\it Nucl. Phys.}
{\bf B335} (1990) 155.}
The second category of conceptual issues consists of certain ``distasteful
allegations", which string theory might cleverly evade.
The first of these is the claim that effects associated with virtual
black holes cause pure quantum states to evolve into mixed
states.${}^{\mixed}$ If this were true, it would mean that the entire
mathematical
framework of quantum mechanics is inadequate. It seems reasonable to
explore whether string theory could avoid allowing pure states to evolve
into mixed states. To the extent that string theory can be consistently
formulated as an S matrix theory, it seems almost inevitable that this
should work out. A second distasteful allegation is that wormhole
contributions to the Euclidean path integral${}^{\worm}$
render the parameters of
particle physics stochastic.${}^{\prob}$
In a previous paper,${}^{\jhscurse}$ I referred to this
phenomenon as `the curse of the wormhole,' since it would imply that
even when the correct microscopic theory is known, it will still not be
possible to compute experimental parameters such as coupling constants,
mass ratios, and mixing angles from first principles.

\REF\hawking{S. W. Hawking, {\it Commun. Math, Phys.} {\bf 43} (1975) 199;
{\it Phys. Rev.} {\bf 13} (1976) 191.}
\REF\zurek{W. H. Zurek and K. S. Thorne, {\it Phys. Rev. Lett.} {\bf 54}
(1985) 2171; K. S. Thorne, R. H. Price, and D. A. Macdonald,
{\it Black Holes: The Membrane Paradigm} (Yale University Press, 1986).}
\REF\thooft{G. `t Hooft, {\it Nucl. Phys.} {\bf B335} (1990) 138; ``The Black
Hole Horizon as a Quantum Surface," {\it Nobel Symposium 79}, June 1990, and
references therein.}
\REF\bghhs{M. J. Bowick, S. B. Giddings, J. A. Harvey, G. T. Horowitz,
and A. Strominger, {\it Phys. Rev. Lett.} {\bf 61} (1988) 2823.}
\REF\preskill{J. Preskill and L. M. Krauss, {\it Nucl. Phys.} {\bf B341}
(1990) 50; J. Preskill, ``Quantum Hair," Caltech preprint
CALT-68-1671, {\it Nobel Symposium 79}, June 1990;
S. Coleman, J. Preskill, and F. Wilczek,
{\it Mod. Phys. Lett.} {\bf A6} (1991) 1631.}
A third issue concerns the classification of black holes.
According to the ``no hair" theorems, in
classical general relativity black holes are fully characterized by mass,
electric charge, and angular momentum. On the other hand, they have a
(large) entropy that is proportional to the area of the event
horizon.${}^{\hawking}$
This amount of entropy corresponds to a number of degrees of
freedom that is roughly what one would get from a
vibrating membrane just above the horizon.${}^{\zurek}$ In fact, `t Hooft
has tried to make sense of such a physical picture, interpreting the
membrane as a string world sheet.${}^{\thooft}$ Alternatively, the
black hole degrees of freedom might be accounted for in string theory
in more subtle ways that utilize possibilities for evading the
classical no-hair theorems by quantum effects.
Recent studies have shown that black holes can have `quantum
hair,' which is observable (in principle) by generalized Bohm--Aharonov
interference measurements.${}^{\bghhs,\preskill}$
Charges that can characterize quantum hair
for black holes are precisely the same ones whose conservation cannot be
destroyed by wormhole effects. Thus the nicest outcome might be for
the correct fundamental theory to provide so many different
types of quantum hair as to produce precisely
the number of degrees of freedom that
is required to account for the entropy of
black holes. In a theory with enough distinct
degrees of freedom to account for
black hole entropy and to protect quantum coherence,
there should be no deleterious effects due to wormholes.

\REF\kalara{S. Kalara and D. V. Nanopoulos, ``String Duality and Black
Holes," preprint CTP-TAMU-14/91, March 1991.}
\REF\jhsdeep{J. H. Schwarz, ``Can String Theory Overcome Deep Problems
in Quantum Gravity?" preprint CALT-68-1728, May 1991.}
A mechanism that has been proposed as an origin for quantum hair
is for a continuous gauge symmetry to break spontaneously leaving a
discrete subgroup unbroken. As we have discussed,
string theory has a large group of discrete symmetries that can be
understood as remnants of spontaneously broken gauge symmetries, namely
the target space dualities. This fact led me to propose
that these are the relevant symmetries for understanding quantum hair
in string theory.${}^{\jhscurse}$
\foot{This idea has been proposed independently in Ref. [\kalara ],
though the emphasis there is on `duality of the S field.'}
However, following further studies and discussions with others, it has become
clear that this suggestion has serious problems, mostly stemming
from the fact that these symmetries are almost all broken for any
particular choice of vacuum configuration.

\REF\abl{T. J. Allen, M. J. Bowick, and A. Lahiri, {\it Phys. Lett.} {\bf
237B} (1990) 47.}
\REF\rohm{R. Rohm and E. Witten, {\it Ann. Phys. (N.Y.)} {\bf 170} (1986) 454.
C. Teitelboim, {\it Phys. Lett.} {\bf 167B} (1986) 69.}
The first proposal for ``quantum hair" of black holes, detectable only
by Bohm--Aharonov-type interference effects, was put forward a few years
ago by Bowick et al.${}^{\bghhs}$ As initially formulated, the analysis
only applied to theories containing a massless `axion.' However,
a subsequent paper demonstrated that this restriction was not essential
and that a suitable massive axion could do the same job.${}^{\abl}$

Stripping away all interactions, the basic idea
can be explained quite simply. Assume four-dimensional space-time and
let $A_{\mu}$ be a $U(1)$ vector field
and $B_{\mu \nu}$ an antisymmetric tensor gauge field (called the axion).
In the language of forms, the associated field strengths are given by
$H=dB$ and $F=dA$. The action consists of the usual kinetic terms,
schematically given by $S_{kin}= {1\over 2}\int d^4x (H^2+F^2)$, and a
topological mass term of the form $S_{mass}=m\int B\wedge F$.
The equations of motion, $d*H=mF$ and $d*F=-mH$, imply
that both fields have mass $m$. What happens is that the
antisymmetric tensor eats the vector to become massive. In four dimensions
it is equivalent to say that the vector eats the scalar (which is dual
to $B_{\mu\nu}$) to become massive.

The axion charge in a region of three-dimensional space $V$ with
boundary $\partial V$ is defined by
$$Q_{axion}=\int_V H= \int _{\partial V} B. $$
For a space-time with nontrivial second homology, such as
Schwarzschild space-time (whose topology is $S^2\times R^2$), it
is possible to obtain nonzero axion charge while having the $H$ field
vanish outside some central region. In this case the $B$ field on the
enclosing two-surface is proportional to a two-form that is closed but
not exact (i.e., belongs to the second cohomology group). In a theory
with axions there are strings (cosmic or fundamental) that contribute a
term to the action proportional to $\int_{\Sigma} B_{\mu\nu} dx^{\mu}\wedge
dx^{\nu}$. As a result, a world sheet enclosing a black hole with
axionic charge gives a Bohm--Aharonov phase $\exp [2\pi i Q_{axion}]$.
This makes the charge observable through interference effects (modulo
unity). If there were nontrivial third homology, the axion charge itself
would be quantized and nothing would be observable.${}^{\rohm}$
However, this is not the case for a Schwarzschild black hole.

\REF\gands{M. B. Green and J. H. Schwarz, {\it Phys. Lett.} {\bf 149B} (1984)
117.}
In the string theory context, the formula for the field strength $H$ is
embellished by various Chern--Simons terms that were omitted in the
discussion above. Also, one-loop effects in
ten dimensions give contributions to the effective action of the form
$\int{ B\wedge tr(F^4)}$, which play a crucial role in anomaly
cancellation.${}^{\gands}$ Upon compactification it can happen that
the ten-dimensional gauge fields acquire expectation values that
result in a nonvanishing effective term of the form $\int B\wedge F$,
where $F$ is a $U(1)$ gauge field in four dimensions, as required to give mass
to the axion. This happens when the associated $U(1)$ gauge symmetry in
four dimensions appears to be anomalous by the usual criteria based on
triangle diagrams. However, as in ten dimensions, $B_{\mu\nu}$ has
nontrivial gauge transformation
properties that give compensating contributions and render
the quantum theory consistent.

Axion charge appears to be a good candidate
for quantum hair in string theory. Of course, if this particular charge were
the only type of quantum hair in string theory, we would still be
very far from achieving the goal of finding enough quantum degrees of
freedom to account for all the entropy of black holes and overcoming
the other problems in quantum gravity that we have discussed.

Fortunately, string theory seems to allow various
categories of generalizations of axion charge that could provide many more
kinds of quantum hair. For example,
the field $B_{\mu\nu}(x,y)$ is defined in ten dimensions. (Here $x$
refers to four-dimensional space-time and $y$ to six compactified
dimensions.) In the usual Kaluza--Klein fashion,
this represents an infinite family of four-dimensional
fields $B_{\mu\nu}^{(n)}(x)$ corresponding to an expansion in harmonics
of the compact space of the form $\sum C_n(y) B_{\mu\nu}^{(n)}(x)$. The
analysis above only utilized the axion corresponding to the leading term
in this series for which $C(y)$ is a constant. The other terms describe
fields that naturally have masses of the order of the compactification
scale. It seems plausible that they could provide additional
types of quantum hair. (When the analysis is done carefully,
target-space duality may yet prove to be important!)
Even this infinite collection of charges may not be the end of the story. The
massive string spectrum contains an infinite number of gauge fields of every
possible tensor structure. The particular gauge field $B_{\mu\nu}$
is special by virtue of its coupling to the string world sheet, which
played a crucial role in the  reasoning above. Other gauge fields
enter the world sheet action with couplings given by their associated
vertex operators. For fields that are not massless in ten dimensions
these give nonrenormalizable couplings in the sigma model, and are
therefore difficult to analyze. Still, from a more general string field
theory point of view, they are not really very different, and so there
may be many more possibilities for quantum hair
associated with the massive string spectrum.

\REF\garfinkle{D. Garfinkle, G. T. Horowitz, and A. Strominger,
{\it Phys. Rev.} {\bf D43} (1991) 3140; see
also G. Gibbons, {\it Nucl. Phys.} {\bf B207} (1982) 337 and G. Gibbons and A.
Maeda, {\it Nucl. Phys.} {\bf B298} (1988) 741.}
\REF\patricia{ J. Preskill, P. Schwarz, A. Shapere, S. Trivedi, and F.
Wilczek, ``Limitation on the Statistical Description of Black Holes,"
preprint IASSNS-HEP-91-34.}

In addition to string symmetries altering some consequences of
general relativity at the quantum level, it is also possible that
special features of string theory play an important role at the
classical level. One indication of this appears in a recent study
of charged black holes,${}^{\garfinkle}$ where
effects of the dilaton field make qualitative changes
already at zeroth order in $\alpha'$. Specifically, whereas a
Reissner--Nordstrom black hole of mass $M$ and charge $Q$ has its
horizon at the radius $R_H=M+\sqrt{M^2-Q^2}$, the corresponding string
solution has $R_H=2M\sqrt{1-Q^2/2M^2}$. Also, as the charge of the black
hole approaches its maximum allowed value, the entropy $S\to 4\pi M^2$
and the temperature $T\to 0$ in the Reissner--Nordstrom case. In the
string case one finds $S\to 0$ and $T\to 8\pi M$. (All these results
are to leading order in $\alpha'$ and $\hbar$.) As one might expect, the
thermodynamic description breaks down in either of these extreme
limits.${}^{\patricia}$

\medskip
\noindent{\bf 5. The Importance of Supersymmetry}

In order to gain the attention and
respect of our experimental colleagues it is
important to make predictions that bear
on near-term experimental possibilities. It is pretty clear that the
best prospect in this regard is supersymmetry. It would be nice if we
could honestly assert that supersymmetry (broken at the weak scale) is
an inevitable feature of any quasi-realistic string model and thus a
necessity if string theory is the correct basis of unification.
Experimentalists
could then be in a position to demonstrate that ``string theory is
false" or to discover important evidence in its support. But can we
honestly make such an assertion?

Certainly no string models that are remotely realistic have been
constructed without low-energy supersymmetry. Also, in the context of
string theory any alternative mechanism for dealing with the
hierarchy problem seems very unlikely. Still, if as is generally
assumed, string theory has a clever way of preventing a
cosmological constant from arising once supersymmetry is broken, then
maybe it could also have a clever stringy alternative for preventing
Higgs particles from acquiring unification scale masses through
radiative corrections. No plausible alternative to supersymmetry
is known, but how sure can we be that one doesn't exist? There is
always the possibility that some mechanism, not yet considered, could be
important, but that shouldn't completely prevent from us ever sticking our
necks out a bit. I don't think it would be dishonest for string theorists
to assert that according to our present understanding, supersymmetry
broken at the weak scale is required by string theory.

\REF\susygut{J. Ellis, S. Kelley, and D. V. Nanopoulos, {\it Phys. Lett.} {\bf
249B} (1990) 441 and {\bf 260B} (1991) 131; U. Amaldi, W. de Boer, and
H. F\"urstenau, {\it Phys. Lett.} {\bf 260B} (1991) 447;
P. Langacker and M.-X. Luo,
preprint UPR-0466T (1991).}

In fact, the experimental prospects are beginning to look up. The
requirement that the three couplings of the standard model should
become equal at a unification scale fails badly without supersymmetry.
On the other hand, for a susy scale ranging from 100 GeV to 10 TeV they
merge very nicely at about $10^{16}$ GeV.${}^{\susygut}$
While this is far from
conclusive, it is a very impressive bit of evidence. Given the
present experimental situation, together with various theoretical and
astrophysical considerations, it seems quite plausible that the
lightest supersymmetry particles are at the low end of this range. Others,
such as squarks and gluinos, maybe be around a TeV or so. If this is
correct, it is unlikely that any of these particles will be produced
and detected before the LHC or SSC comes on line. However, supersymmetry
has important implications for the Higgs sector that could be confirmed
sooner.

Low-energy supersymmetry has two Higgs doublets, which after symmetry
breaking result in a charged particle $H^{\pm}$, and three neutral
particles $h$, $H$, and $A$. The minimal supersymmetric standard model
(MSSM) requires, at tree level, that $h$ is the lightest of these
and that its mass not exceed $M_{Z}$. When radiative corrections are taken into
account, it can be somewhat heavier, depending on the mass of the
top quark. For example, if the top quark mass does not exceed 160 GeV then the
$h$ mass should not exceed 120 GeV. For a top quark mass below 130 GeV
the bound is lowered to 100 GeV. These bounds need not be saturated, so
there is a reasonable chance for a mass in the range 50-100 GeV, making
it open to discovery at LEP 2. Another interesting possibility is that
the top quark could decay into $H^{+}$ plus a bottom quark. If it is
kinematically allowed, this could be a significant branching fraction.
(The precise prediction depends on the parameter $tan \beta = v_2/v_1$.)
I am optimistic that some of these particles will turn up during this
decade and that this will open up an exciting era for string theorists (as well
as all
particle physicists).

\singlespace
\refout

\end